\documentclass[twocolumn,aps,prb,superscriptaddress,floatfix]{revtex4-2}
\usepackage{lineno}

\usepackage{xcolor}
\usepackage{graphicx}
\usepackage{dcolumn}
\usepackage{bm}
\usepackage{amsmath}

\usepackage[T1]{fontenc}
\usepackage[unicode=true,pdfusetitle,
 bookmarks=true,bookmarksnumbered=false,bookmarksopen=false,
 breaklinks=false,pdfborder={0 0 1},backref=false,colorlinks=false]
 {hyperref}
\hypersetup{
 colorlinks,linkcolor=blue,citecolor=blue,urlcolor=blue}

\newcommand{\tep}{\tilde{\varepsilon}}
\newcommand{\tH}{\tilde{H}}

\newcommand{\fig}[1]{FIG. #1}
\newcommand{\subfig}[2]{\fig{#1} (#2)}

\begin{document}


\title{\textcolor{black}{Orbital Hall Responses in Disordered Topological Materials}}

\author{Luis M. Canonico}
\email{luis.canonico@icn2.cat}
\affiliation{Catalan Institute of Nanoscience and Nanotechnology (ICN2), CSIC and BIST, Campus UAB, Bellaterra, 08193 Barcelona, Spain}
\author{Jose H. Garcia}
\
\affiliation{Catalan Institute of Nanoscience and Nanotechnology (ICN2), CSIC and BIST, Campus UAB, Bellaterra, 08193 Barcelona, Spain}
\author{Stephan Roche}
\affiliation{Catalan Institute of Nanoscience and Nanotechnology (ICN2), CSIC and BIST, Campus UAB, Bellaterra, 08193 Barcelona, Spain}
\affiliation{ICREA, Instituci\'o Catalana de Recerca i Estudis Avançats, 08070 Barcelona, Spain}

\date{\today}

\begin{abstract}
We report an efficient numerical approach to compute the different components of the orbital Hall responses in disordered topological materials from the Berry phase theory of magnetization. The theoretical framework is based on the Chebyshev expansion of Green's functions and the {\color{black}off-diagonal} elements of the position operator for systems under arbitrary boundary conditions. The capability of this scheme is shown by computing the orbital Hall conductivity for gapped graphene and for Haldane model in presence of nonperturbative disorder effects. This methodology enables realistic simulations of orbital Hall responses in highly complex models of disordered materials.
\end{abstract}

\maketitle

Similarly to spintronics, \emph{orbitronics} operate on stimuli capable of generating nonequilibrium distributions of orbital angular momentum\cite{Go-Review,atencia2024orbital}. Seminal works from Bernevig \textit{et al.} and Kontani \textit{et al.} predicted the existence of an orbital Hall effect (OHE) in doped semiconductors\cite{OHEBernevig} and transition metals\cite{OHEorigins-3}, where a longitudinal current triggers a transverse flow of orbital angular momentum (OAM) independently of the spin-orbit coupling (SOC) of the material. Initially, the potential of the orbital degrees of freedom for technological applications has been questioned under the assumption of the quenching of the orbital angular momentum due to the crystal field\cite{bihlmayer2022rashba}. However, recently, Go \textit{et al.} demonstrated that even in systems with quenched orbital character, the application of electric fields enables hybridization channels absent at equilibrium, giving rise to the OHE\cite{OrbitalTexture}. On the other hand, the experimental detection of the OHE was first investigated indirectly through orbital torque measurements\cite{orbital-torque-magnetic-bilayers-EXP,hayashi2023observation,fukunaga2023orbital,Orbital-torque-1,bose2023detection,Gambardella}, orbital pumping\cite{SantosOPumping,go2023OrbitalPumping}, and inverse OHE\cite{seifert2023time,xu2024orbitronics}. In these studies, the validity of the OHE scenario remains however questioned due to the similarities in the symmetries obeyed by the OHE and the spin Hall effect, but two independent experiments have supported the existence of the OHE through magneto-optical Kerr rotation measurements in Ti\cite{Exp-OHE-1} and Cr\cite{ExpOHE2}. 

To date, most of those experimental results on the electrical generation of orbital currents rely on three-dimensional (3D) systems. Nonetheless, the tunability in the properties of two-dimensional (2D) materials and the prospect of developing ultra-compact light-metal-based orbitronic devices has gained significant attention. For instance, theoretical works predicted that 1H transition metal dichalcogenides (TMDs) could host orbital-Hall insulating phases\cite{OHE_Bhowal_1,Us2} characterized by an orbital Chern number and in-gap OAM-carrying edge states\cite{Us3}. Moreover, sizable OHE was predicted in semiconducting 1T TMDs\cite{costa2022connecting} and phosphorene \cite{cysne2023ultrathin}. Finally, the role of the orbital degrees of freedom in generating real-space localized non-equilibrium spin densities and their suitability for SOT was recently revealed \cite{canonico2023spin}.

Beyond the muffin-tin picture that considers the contributions to the OAM from the immediacies of the atomic environment encoded in the orbital character of the wavefunctions, nonlocal contributions arising from the electron wavepacket self-rotation {\color{black} have proven to be challenging to estimate due to the difficulties in representing the position operator in periodic systems. Methodological developments addressed it using the Berry phase \cite{adams1959energy,blount1962formalisms,RestaBianco}, leading to the modern theory of orbital magnetization \cite{BerryPhaseElProps,OrbitalPeriodic,resta2005orbital,thonhauser2011theory,bianco2016orbital, PRoleBerryMoment}, and applications have been studied in gapped\cite{OHE_Bhowal-Vignale,nonlocalMario}, bilayer\cite{TunableReversableMagnetoelectricBG,cysne2024controlling} and twisted bilayer graphene\cite{he2020giant,serlin2020intrinsic}, as well as gapped graphene nanoribbons \cite{kazantsev2024nonconservation} and kagome lattices\cite{BuschMertig}.}

However, despite all the theoretical and experimental development, practical applications of orbitronics still require a better quantification and understanding of the role of the disorder in the generation of orbital currents and the relaxation of nonequilibrium orbital densities. Examples of this necessity are the results from the experiments of Seifert \textit{et al.} in magnetic bilayers, where magnetic disorder suppresses nonequilibrium spin densities and magnonic contributions, leaving their orbital currents and densities unaffected, which help in identifying a ballistic conduction of OAM \cite{seifert2023time}. On the other hand, Choi \textit{et al.} reported an orbital diffusion length $l_O$, in Ti between $61-74$ nm\cite{Exp-OHE-1} that contrasts with the short orbital diffusion length of $l_O=6.6$ nm estimated by Lyalin \textit{et al.} in Cr\cite{ExpOHE2}. Such variation of $l_O$ might be related to crystal field or disorder effects which remain mostly unexplored. The work of Pezo \textit{et al.} suggests that random impurities decrease the value of the orbital Hall conductivity in Dirac materials \cite{PezoDiracMaterials}, while other {\color{black} studies} report that weak disorder can be the dominant source of current-induced orbital currents in the valence and conduction bands of Dirac materials\cite{liuculcer2023dominance,veneri2024extrinsicorbitalhalleffect}. Conversely, for triangular lattices, the disorder appears to be detrimental\cite{tanggerrit2024role}. In that context, the computational cost of diagonalization methods to treat disorder effects becomes quickly prohibitive, and there is a need for realistic, accurate, and efficient simulation tools capable of addressing non-perturbatively the role of the disorder in the generation of orbital currents in systems approaching experimentally relevant scales.

In this letter, we present an efficient linear-scaling method that allows the computation of the orbital Hall conductivity (OHC) in disordered materials. Using the kernel polynomial method (KPM), we expand the position operator, the OAM, and the Kubo-Bastin formula. We illustrate the methodology by studying the transport properties of orbital angular momentum in the topologically trivial and non-trivial phases of the Haldane model in presence of Anderson disorder. Our results show that disorder favors the appearance of extrinsic {\color{black}Fermi-surface} contributions that overcome the intrinsic orbital Hall responses for the clean case. The generality of the methodology enables the study of various current-induced orbital responses in disordered systems in a straightforward manner, independently of a basis choice for any temperature and chemical potential.

To address the components of the conductivity tensor, we use the linear response Kubo formalism. In the static limit, the elements of the OHC tensor are given by the Kubo-Bastin formula \cite{Us1,bastin1971quantum,Us2}
{\color{black}
\begin{align}
\sigma_{\alpha\beta}^{k} = \frac{i 2\hbar e}{\Omega}\int_{-\infty}^{\infty}d\varepsilon F(\varepsilon)
\text{Im}\text{Tr}\langle J_{\alpha}^{k}\partial_{\varepsilon}G^{+}(\varepsilon,H)v_{\beta}\delta(\varepsilon - H)\rangle,\label{eq:bastin}
\end{align}}

\noindent where $\Omega$ is the volume of the sample, $v_{\beta}$ is the $\beta$ component of the velocity operator $v_{\beta}\equiv \frac{i}{\hbar}[H,r_{\beta}]$, {\color{black}$ G^{\pm}(\varepsilon,H)=\frac{1}{\varepsilon -H \pm i0^{+}}$} is the {\color{black}retarded (advanced)} Green's function, {\color{black}$F(\varepsilon)=(\exp((\varepsilon-\mu)/K_BT)+1)^{-1}$} is the Fermi-Dirac distribution for a given temperature $T$ and chemical potential $\mu$, {\color{black}$J_{\alpha}^{k}=\frac{1}{2}\lbrace L_{k},v_{\alpha} \rbrace$} is the orbital current operator. Though the eigenstate-based representation of the Kubo formula has been extensively used in the study of orbital angular momentum transport in clean systems \cite{Go-Review,OHE_Bhowal-Vignale,Us3}, equation \eqref{eq:bastin} has been instrumental in the investigation of electrical responses in disordered materials \cite{Luis-PRL,Fan2021linear,kite}. Following the approach proposed by Bhowal and Vignale\cite{OHE_Bhowal-Vignale}, we write the symmetrized orbital angular momentum operator as $\vec{L}=\frac{e\hbar}{4g_L\mu_B}(\vec{r}\times \vec{v}- \vec{v}\times\vec{r})$, where $g_L$ is the Land\'e $g$-factor for the orbital angular momentum, and $\mu_B$ is the Bohr magneton. From this definition, it is clear that the difficulties in computing the orbital angular momentum beyond the muffin-tin approximation are related to the ill-defined nature of the position operator in periodic systems. Nonetheless, as noted by Bianco and Resta \cite{RestaBianco}, even if the diagonal elements in the energy eigenstates of the position operator are ill-defined, the off-diagonal elements can be easily written as $\langle i|r_{\alpha}|j\rangle = i\hbar\frac{\langle i| v_{\alpha}|j\rangle}{E_j - E_i}$, {\color{black}where $|i\rangle$ and $|j\rangle$ are eigenstates of the Hamiltonian $H$ with energies given by $H|i\rangle=E_i|i\rangle$ and $H|j\rangle=E_j|j\rangle$.} Thus, using the definition of Green's functions,  we define equivalent representations of the same position operator as

{\color{black}
\begin{align}
\langle i|r_{\alpha}^{+}|j\rangle =i\hbar\langle i|\int d\varepsilon^{\prime}\left[\text{Re}(G^{+}(\varepsilon^{\prime},H))v_{\alpha}\delta(H-\varepsilon^{\prime})\right]|j\rangle\nonumber&&\\
\langle i|r_{\alpha}^{-}|j\rangle  =-i\hbar\langle i|\int d\varepsilon^{\prime}\left[\delta(H-\varepsilon^{\prime})v_{\alpha}\text{Re}(G^{-}(\varepsilon^{\prime},H))\right]|j\rangle,\label{eq:posop}
\end{align}
}

\noindent with {\color{black} $\text{Re}(G^{\pm}(\varepsilon^{\prime},H)) = \frac{1}{2}(G^{+}(\varepsilon^{\prime},H)+G^{-}(\varepsilon^{\prime},H))$.} These representations of the $r_{\alpha}$ operator in \eqref{eq:posop} are formally equivalent in the limit of vanishing broadening. However, for their numerical evaluation, both versions of the operator must be considered to cancel any spurious contributions due to the regularization of Green's functions. The numerical evaluation of equation \eqref{eq:posop} is inefficient if one uses the eigenstates of the system, in contrast with spectral methods such as the KPM which perform computation as matrix-vector operations. Following the Chebyshev expansion technique\cite{WeisseKPM,Fan2021linear}, we rescale the Hamiltonian $H$ and the energies $\varepsilon$ between the $\left(-1,1\right)$ interval and expand the Green's and spectral functions as a polynomial series of the rescaled Hamiltonian $\tilde{H}$ and energies $\tilde{\varepsilon}$. Thus the expanded expression for {\color{black}$r^{+}_{\alpha}$} in \eqref{eq:posop} is 
{
\color{black}
\begin{align}
\langle i|r_{\alpha}^{+}|j\rangle = i\hbar\frac{2}{\Delta E}\int_{-1}^{1} d\tilde{\varepsilon}^{\prime}\langle i |\left(\sum_{\mu=0}^{M}\text{Re}(c_{\mu}(\tilde{\varepsilon}^{\prime}))T_{\mu}(\tH)\right) v_{\alpha}&&\nonumber\\
\times\left(\sum_{\nu=0}^{M}f_{\nu}(\tilde{\varepsilon}^{\prime})T_{\nu}(\tH)\right)|j\rangle.
\end{align}
}
The coefficients {\color{black} $c_{\mu}(\tep)=\frac{-2i}{\sqrt{1-\tep^2}}\frac{g_{\mu}e^{-i\mu\arccos(\tep)}}{(\delta_{\mu,0}+1)}$ }and $f_{\nu}(\tep)=\frac{2}{\pi \sqrt{1-\tep^2}}\frac{g_{\nu}T_{\nu}(\tep)}{(\delta_{\mu,0}+1)}$ are related to the polynomial expansion of $G^{+}(\tep,\tH)$ and $\delta(\tH-\tep)$, respectively\cite{GarciaRappoportConductivity}, $\Delta E$ is the energy width of the spectrum defined as $ \Delta E = (E_{max}-E_{min})/2$, $T_{m}(x)=\cos(m\arccos(x))$ is the $m-$th Chebyshev polynomial of first-kind and $g_\mu$ is a damping factor added to the series to smooth the Gibbs oscillations resulting from the truncation of the polynomial expansion at order $M$\cite{WeisseKPM} {\color{black}(See SM for more detailed presentation of the KPM\cite{SuplementaryMaterial})}. The components of {\color{black}$r^{-}_{\alpha}$} are obtained from the relation {\color{black}$r_{\alpha}^{-}=\left(r_{\alpha}^{+}\right)^{\dagger}$}. Using these expansions, the orbital angular momentum operator $L_k$ is written as

\begin{figure}[h]
\centering
\includegraphics[width=.9\linewidth]{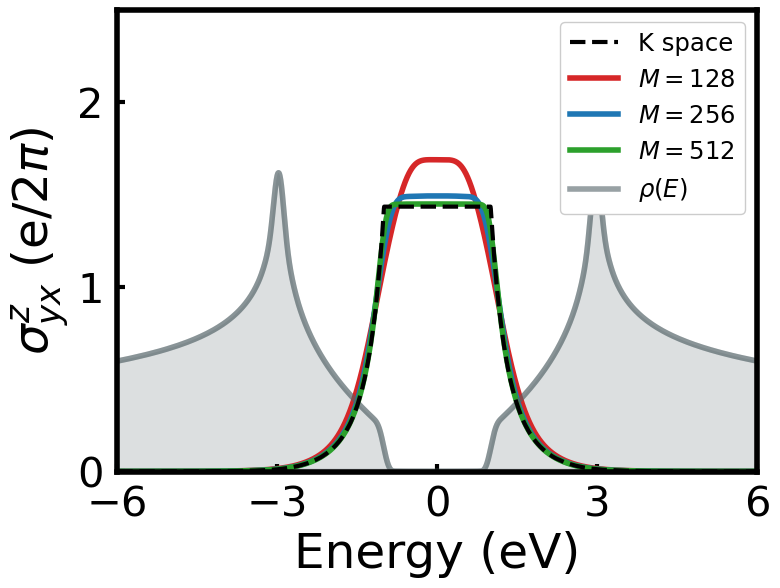}
\caption{Real-space orbital Hall conductivity of the Haldane model with $t_2=0$, $\Delta=1.0$ eV, and $W=0$ (for different number of moments $M$, and compared with the diagonalization result (dashed line)). The shaded curve represents the DOS computed for $M$=512.}
\label{fig:fig1}
\end{figure}

{
\color{black}
\begin{equation}
    L_{k}=\epsilon_{ijk}\frac{e\hbar^{2}}{4g_L\mu_B}(\frac{r_{i}^{+}}{\hbar}v_{j}- v_{i}\frac{r_{j}^{-}}{\hbar})=\frac{e\hbar^2}{4g_L\mu_B}\ell_{k},
\end{equation}
}
\noindent with $\epsilon_{ijk}$, being the Levi-Civita symbol, and defined {\color{black}$\ell_k=\epsilon_{ijk}(\frac{r_i^{+}}{\hbar}v_j - v_i \frac{r_j^{-}}{\hbar})$}. Inserting the expansion of the $L_k$ operator into \eqref{eq:bastin} and using the polynomial expansion of the Green and spectral functions, the components of the OHC tensor reads as: 
{\color{black}
\begin{align}
    \sigma_{\alpha\beta}^{k} &= -\frac{8}{{\Delta E}^3} \frac{4}{\pi} \frac{e^2 \hbar^3}{4g_L\mu_B\Omega} \nonumber&&\\
    {}&\times\int_{-1}^{1}d\tep \frac{F(\tep)}{(1-\tep^2)^2}\sum_{m,n} 2 \mu_{m,n}^{\alpha,\beta,k}\text{Im}\left(\Gamma_{m,n}(\tep)\right),\label{eqn:fexpansion}
\end{align}
}
with the expansion coefficients defined as 

{\color{black}
\begin{equation}
    \mu_{m,n}^{\alpha,\beta,k}=\frac{g_m g_n}{\left(1+\delta_{m,0}\right)\left(1+\delta_{n,0}\right)}Tr\langle \frac{1}{2}\lbrace\ell_{k},v_{\alpha}\rbrace T_m(\tilde{H})v_{\beta}T_n(\tilde{H})\rangle.\label{eqn:moments}
\end{equation}}
\noindent Here, the coefficients in \eqref{eqn:moments} contain all the information related to the Hamiltonian, the orbital current operator{\color{black}, and the different disorder effects in the system}. It neither depends on the energy nor the specific basis chosen for describing the system and comprises the most expensive part of the calculation (details on the computation of these coefficients are shown in SM\cite{SuplementaryMaterial}). In contrast, $\Gamma_{m,n}(\tep)$ {\color{black} reads as}
{\color{black}

\begin{equation}
\Gamma_{m,n}(\tep)=\left(-i \tep+  m \sqrt{1-\tep^2}\right) e^{-im\arccos(\tep) }T_n(\tep),
\end{equation}}
\noindent {\color{black} it serves as the generalized basis of the expansion, is independent of the Hamiltonian and its  integration in \eqref{eqn:fexpansion} accounts for all the energy-allowed processes' contributions to the total orbital-Hall conductivity.} Equation \eqref{eqn:fexpansion} constitutes the main result of our work. After determining the coefficient matrix {\color{black}$\mu_{m,n}^{\alpha,k}$}, we can compute the energy-resolved OHC for all the temperatures. To evaluate the traces, we use the random phase approximation which allows the expansion of the trace in terms of a subset of elements\cite{WeisseKPM} {\color{black}(See SM for more details\cite{SuplementaryMaterial})}.

\begin{figure}[h]
\centering
\includegraphics[width=0.99\linewidth]{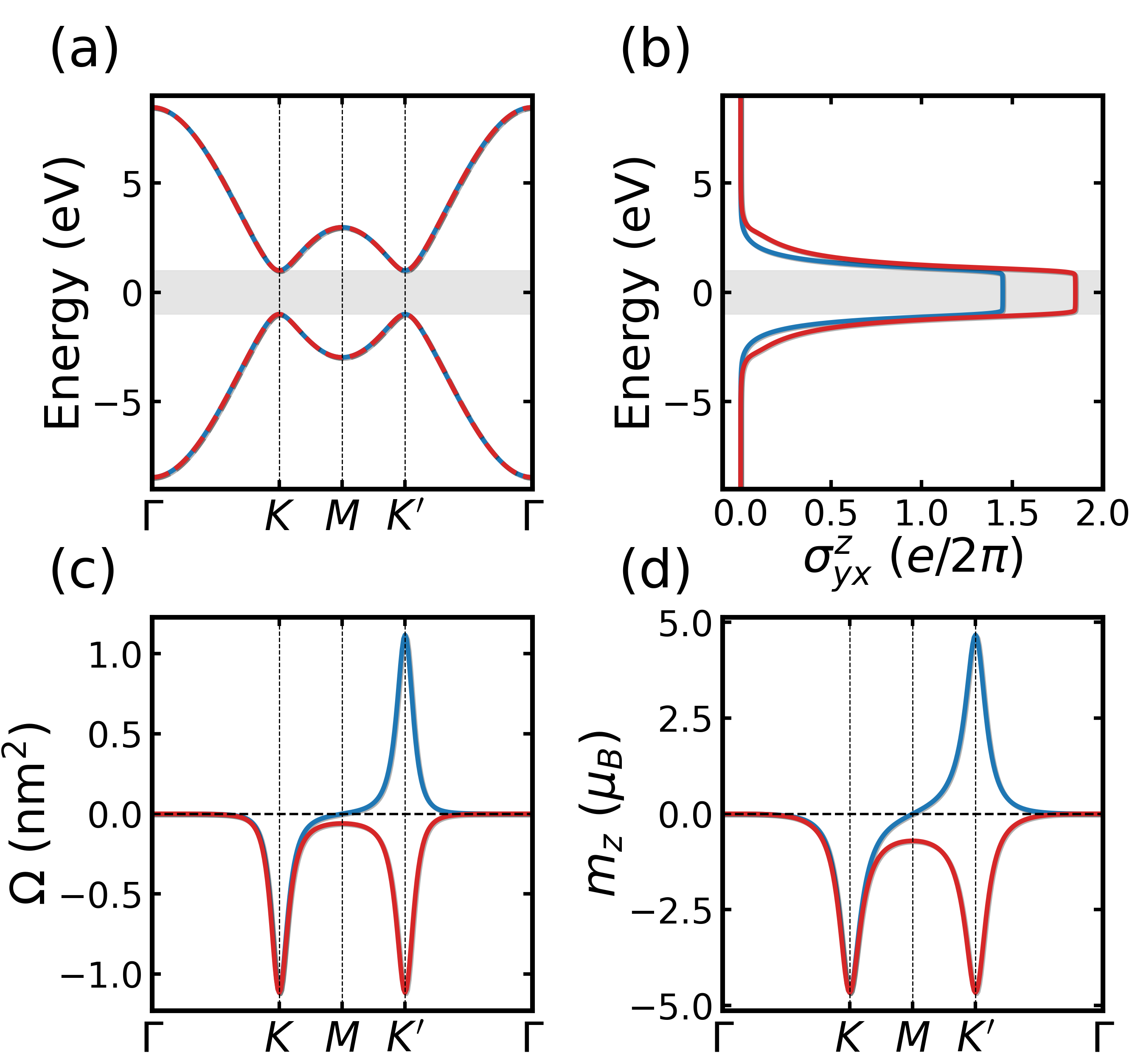}
\caption{(a) Energy bands of the Haldane model in the trivial (blue lines) phase with $\Delta=1.0$ eV and $t_2=0$, and the nontrivial (red lines) phase $t_2=1.0$ eV and $\Delta=0$. (b) Orbital Hall conductivity for the Haldane model in the trivial and nontrivial phases, computed for systems with {\color{black}$512\times512$} unit cells and $M=512$. (c) Berry curvature of the lowest energy band of the Haldane model in the trivial and nontrivial topological phases. (d) Orbital magnetic moment for the lowest energy band of the Haldane model in both phases.} \label{fig:fig2}
\end{figure}

For illustration, we apply this method to study the transport of orbital currents in the Haldane model\cite{HaldaneModel}. The Hamiltonian is given as

{\color{black}
\begin{equation}
H=-t\sum_{\langle i,j\rangle }c^{\dagger}_i c_j + i\frac{t_2}{3\sqrt{3}}\sum_{\langle \langle i,j \rangle \rangle}\nu_{ij} c^{\dagger}_i c_j+\sum_{i}(\Delta_i +\xi_i)c^{\dagger}_i c_i,
\end{equation}}

\begin{figure*}[htp]
\centering
\includegraphics[width=0.99\linewidth]{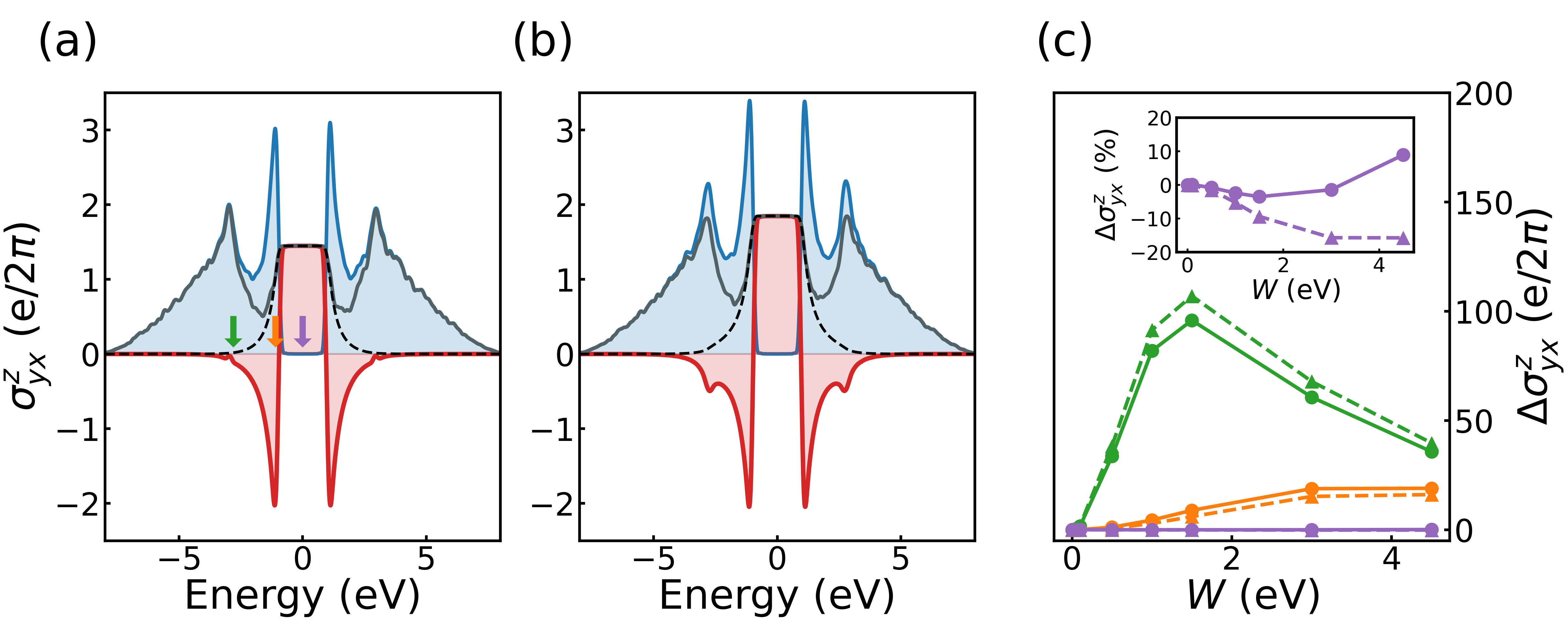}
\caption{Total (grey line), Fermi sea (red {\color{black} shaded area}) and Fermi surface (blue shaded area) contributions of the orbital Hall conductivity for the Haldane model in the trivial phase for $t_2=0$, $\Delta=1.0$ eV, and $W=0.1$ eV (a), and the topologically nontrivial phase for $t_2=1.0$ eV, $\Delta=0.0$, and $W=0.1$ eV (b) computed for systems with {\color{black}$512\times 512$} unit cells, $M=512$  and averaged over {\color{black}$160$} disorder configurations. The case for $W=0.0$ (dashed line) is shown as a guide to the eye. (c) Variation of the orbital Hall conductivity with $W$ for the three energies highlighted in panel (a), the solid and dashed lines correspond to the trivial and topologically nontrivial phases of the Haldane model, respectively. Inset: Percentual variation of the orbital Hall conductivity plateau with respect to the case with $W=0$ for the topologically trivial and nontrivial phases.}
\label{fig:fig3}
\end{figure*}

\noindent where $t$ and $t_2$ are the first and second nearest neighbor hoppings, respectively. $\nu_{ij}=+1(-1)$ if the second nearest neighbor hopping from the sites $j$ to $i$ occurs in the counter-clockwise (clockwise) direction. {\color{black} The $\langle ...\rangle $ and $\langle\langle...\rangle\rangle$ summations run over the first and second nearest neighbor sites}, $\Delta_i=\Delta (-\Delta)$, is the onsite energy for $i\in A (i\in B)$ sublattice, and $\xi_{i}$ is an on-site Anderson disorder term whose values are selected from a uniform random distribution in the interval $[-W/2,W/2]$. Following Ref. \onlinecite{OHE_Bhowal-Vignale}, we fix $t=2.8$ eV and the {\color{black}nearest neighbor distance} $a=1.42$ \AA.

Our starting point is the study of the convergence of the OHC plateau for various values of $M$. Figure \ref{fig:fig1} shows the OHC obtained using the reciprocal space procedure described in Ref.\onlinecite{OHE_Bhowal-Vignale} and our real-space method for clean gapped graphene ($t_2=0$). {\color{black} We considered real-space lattices composed of $512\times512$} unit cells with an onsite potential $\Delta=1$ eV, using the Jackson kernel {\color{black} to damp the Gibbs oscilations\cite{WeisseKPM}}. As the figure shows, for smaller values of $M$ the OHC plateau is broadened and it goes to a slightly higher value than the exact (reciprocal-space computed) OHC plateau. Nonetheless, as $M$ increases the agreement between the two methods increases remarkably. The improvement in the description is due to the reduction of the numerical broadening associated with the Chebyshev polynomial expansion, which for the Jackson kernel decreases as $\sim 1/M$\cite{WeisseKPM,Fan2021linear}.

We then study the interplay between nontrivial topology and the transport of orbital currents. For the sake of comparison, we set the energy gaps of the trivial and nontrivial phases to have the same value (see \subfig{\ref{fig:fig2}}{a}) and compute their OHC. As seen in  \subfig{\ref{fig:fig2}}{b}, for both cases, the OHC does not exhibit quantized values due to the nonconserved nature of the orbital angular momentum, with the topologically nontrivial case having a slightly larger OHC than the trivial phase. As Refs. \onlinecite{OHE_Bhowal-Vignale,Cysne-Bhowal-Vignale-Rappoport,pezo2022orbital,BuschMertig} pointed out, the transport of the orbital angular momentum in systems with quenched orbital character occurs solely due to the combined action of the Berry curvature and the orbital moment distribution in reciprocal space. Thus, it is clear that the simultaneous inversion of the Berry curvature and orbital moment that occurs in the trivial phase due to the time-reversal symmetry (see \subfig{\ref{fig:fig2}}{c} and \subfig{\ref{fig:fig2}}{d}) leads to an overall smaller OHC for the topologically trivial case. These results highlight the dependence of the OHC on the geometric features of the states of the system through the Brillouin zone and add up to the prior theoretical evidence showing that in topologically nontrivial phases, such as the quantum spin-Hall and quantum anomalous-Hall phases, the topologically protected edge states can carry finite orbital angular momentum \cite{Us1}.

We further study the effects of Anderson disorder on the OHC in both trivial and topologically non-trivial phases of the Haldane model. {\color{black} To highlight the modifications to the OHC induced by the disorder, we use the Streda and Smrcka decomposition of the Kubo formula\cite{stvreda1975galvanomagnetic} to isolate the in-gap contributions from the contributions at the Fermi surface to the total OHC.} \subfig{\ref{fig:fig3}}{a} and \subfig{\ref{fig:fig3}}{b} show the OHC for the trivial and topological phases, respectively, for $W=0.1$ eV (other values of $W$ are studied in SM \cite{SuplementaryMaterial}), together with the clean case as a guide to the eye (dashed lines). It is seen that the OHC plateaus remain unchanged by the disorder. Aside from this, outside the energy gap, peaks in the OHC emerge in both the Fermi surface and sea contributions and reach their upper value at energies comparable to $t$.  To inquire about these disorder-enabled effects, we track the evolution of the OHC as a function of $W$ for three energies  \subfig{\ref{fig:fig3}}{a}, corresponding to the middle of the energy gap (purple arrow), $100$ meV below the lowest gap edge (orange arrow) and $-2.8$ eV (green arrow). \subfig{\ref{fig:fig3}}{c} shows the changes in the OHC at these energies, where the solid and dashed lines correspond to the trivial and the topologically nontrivial phases, respectively. Contributions at $E=-2.8$ eV increase monotonically with $W$, and saturate for $W\sim1.5$ eV, where multiple scattering and localization effects dominate and reduce the OHC. {\color{black}In SM, we show that these changes are related to the Fermi surface contributions\cite{SuplementaryMaterial}.} In contrast, the OHC plateau presents small changes with the disorder. The inset illustrates this showing the percental change with respect to the clean case. Here, it shows that for $W$ between $[0,1]$ eV, the change in the height of the plateau is below $5\%$ with the plateau for the topological case decreasing in contrast to the trivial case. Increasing values of $W$ accentuate this tendency, and the OHC for the topological case is reduced by almost $20\%$ of its height, while for the trivial case, it is increased by $10\%$. The permanence and relatively small changes in the OHC agree with the analysis of Bernevig \textit{et al.}\cite{Bernevig2006}. One also notes that our results contrast with those from Pezo \textit{et al.} \cite{PezoDiracMaterials}, which might be due to finite size effects. Regarding the increasing contributions at $E =-2.8$ eV, we argue that they might be related to the formation of disorder-enabled current loops similar to those observed in graphene nanoribbons\cite{CurrentPatternsOrbitalMagnetism} and quantum dots\cite{CurrentVorticesGOmes}.

In conclusion, we have presented a numerical method based on the Chebyshev expansion technique to evaluate the orbital-Hall conductivity tensor components and assess the off-diagonal elements of the position operator whatever the boundary conditions. We have illustrated the method through the analysis of the disordered Haldane models. Our findings evidence that OHC plateaus are robust to disorder, but new orbital current channels are generated by disorder within the bulk of the systems, as previously observed in graphene nanoribbons.

\begin{acknowledgments}
 L.M.C. acknowledges T. G. Rappoport, R. B. Muniz, T. P. Cysne, and A. W. Cummings for fruitful discussions. L.M.C. acknowledges funding from Ministerio de Ciencia e Innovación de Espa\~na under grant No. PID2019-106684GB-I00 / AEI / 10.13039/501100011033, FJC2021-047300-I, financiado por MCIN/AEI/10.13039/501100011033 and the European Union “NextGenerationEU”/PRTR". S.R and J.H.G, acknowledge funding from the FLAG-ERA grant MNEMOSYN, by MCIN/AEI /10.13039/501100011033 and European Union "NextGenerationEU/PRTR”  under grant PCI2021-122035-2A-2a and funding from the European Union’s Horizon 2020 research and innovation programme under grant agreement No 881603. ICN2 is funded by the CERCA Programme/Generalitat de Catalunya and supported by the Severo Ochoa Centres of Excellence programme, Grant CEX2021-001214-S, funded by MCIN/AEI/10.13039.501100011033.  This work is also supported by MICIN with European funds‐NextGenerationEU (PRTR‐C17.I1) and by 2021 SGR 00997, funded by Generalitat de Catalunya.
\end{acknowledgments}

\onecolumngrid

\pagebreak

\section*{Supplementary material for ``Orbital Hall Responses in Disordered Topological Materials''}

\section{Expansion of spectral functions using Chebyshev Polynomials}

We performed our transport calculations using the Kernel Polynomial Method \cite{silver1996kernel,WeisseKPM,Fan2021linear} (KPM).  This procedure consists of the expansion of energy and Hamiltonian functions in powers of the Hamiltonian operator. Here, we present the general method and demonstrate its application in approximating the density of states. First, we must rescale the Hamiltonian $H$ and the energies $\varepsilon$ to the interval $(-1,1)$, where the Chebyshev polynomials are bounded to guarantee the convergence of the series expansion. Thus, the rescaled energies and Hamiltonians are written as $\tilde{\varepsilon}= (E-b)/a$ and $\tilde{H} = (H-b)/a$, respectively. Here $a=(E_{max}-E_{min})/2$ and $b=(E_{max}+E_{min})/2$; $E_{max}$ and $E_{min}$ represent the maximum and minimum energies of the spectrum, respectively. To expand a rescaled spectral quantity $\tilde{f}(\tep,\tH)$, we use the spectral representation

\begin{equation}
	\tilde{f}(\tep,\tH) = \sum_{n}\tilde{f}(\tep,\tilde{E}_n)|\tilde{E}_n\rangle\langle\tilde{E}_n|,
	\label{eqn:spectralrep}
\end{equation}

\noindent where $\tH|\tilde{E}_n\rangle = \tilde{E}_n|\tilde{E}_{n}\rangle$ and expand each of the functions as\cite{WeisseKPM,Fan2021linear}

\begin{equation}
	\tilde{f}(\tep,\tilde{E}_n) = \frac{2}{\pi}\sum_{m=0}^{\infty}\Gamma_m(\tep)T_{m}(\tilde{E}_n),\qquad \Gamma_m(\tep) = \frac{1}{\delta_{m,0}+1}\int_{-1}^{1}d\tilde{E}_n\frac{\tilde{f}(\tep,\tilde{E_n})T_{m}(\tilde{E}_n)}{\sqrt{1-{\tilde{E}_n}^2}},\label{eqn:chebexpcoef}
\end{equation}

\noindent where $T_{m}(\tep)=\cos(m\arccos(\tep))$ are the Chebyshev polynomials of the first kind and can be efficiently computed from the recurrence relation $T_{m+1}(\tep) = 2\tep T_m(\tep) - T_{m-1}(\tep)$, with $T_0 = 1$ and $T_{1}=\tep$. Inserting this expression in \eqref{eqn:spectralrep}, we identify that the spectral expansion of the operator $\tilde{f}(\tep,\tH)$ reads as

\begin{equation}
	\tilde{f}(\tep,\tH) = \frac{2}{\pi}\sum_{m=0}^{\infty}\Gamma_m(\tep)T_m(\tH),
	\label{eqn:chebexpspectral}
\end{equation}
\noindent where now all the information associated with the energy dependence of $\tilde{f}(\tep,\tH)$ has been separated from the information of the Hamiltonian, this expansion technique has been used to approximate various spectral operators, such as Green functions\cite{WeisseKPM,Fan2021linear}, the time-evolution operator\cite{wang1998time,fehske2009numerical} and the density of states\cite{silver1996kernel}. In many cases, only the information contained within the Hamiltonian is needed. An example of this is the density of states $\rho(\varepsilon)$, which depends only on the trace of $H$. Substituting $\tilde{f}(\tep,\tH) = \delta(\tH-\tep)$ in \eqref{eqn:chebexpcoef}, we identify that the density of states is given by

\begin{equation}
	\rho(\tilde{\varepsilon}) = \frac{2}{\pi\sqrt{1-{\tilde{\varepsilon}}^2}}\sum_{m=0}^{\infty}\mu_mT_{m}(\tilde{\varepsilon}),\label{eqn:DOSexpansion}
\end{equation}

\noindent where $\Gamma_m(\tep)= T_{m}(\tep)$ and $\mu_m\equiv \text{Tr}\langle\tH\rangle$. This procedure is general for the expansion of any spectral operator and is exact. However, for practical purposes, it is always necessary to truncate the series at some finite order M. The truncation of the series leads to reduced precision and Gibbs oscillations, especially when the function is not continuously differentiable. Therefore, one can dampen these oscillations and recover the precision of the expansion by modifying the moments $\mu_m\rightarrow\mu_mg_m$ with a kernel $g_m$ that smoothens the high-frequency oscillations in the expansion. In the manuscript, we used the Jackson kernel given by 

\begin{equation}
	g_{m} = \frac{1}{M+1}\left[(M-m+1)\cos\left(\frac{\pi m}{M+1}\right)+ \sin\left(\frac{\pi m}{M+1}\right)\cot\left(\frac{\pi}{M+1}\right)\right].
\end{equation}

This kernel has been recognized as optimal for most situations since it is positive, and normalized, preserving the values of the integration of the expanded function\cite{WeisseKPM}. The evaluation of the expansion moments $\mu_m$ becomes prohibitive for larger systems. However, we can compute them using a stochastic trace evaluation. Thus, instead of performing the whole trace, the coefficients are obtained by

\begin{equation}
	\mu_{m}\approx \frac{1}{R}\sum_{r=0}^{R-1}\langle \psi_r|T_{m}(\tilde{H})|\psi_r\rangle,
\end{equation}

\noindent where $|\psi_r\rangle$ are complex random vectors defined as $|\psi_r\rangle = D^{-1/2}\sum_{i}^{D} \exp(i\theta_i)|i\rangle$. Here, $\lbrace|i\rangle\rbrace_{i=1,...,D}$ denotes the original site and orbital basis set in which the Hamiltonian is represented, and $D$ is the dimension of the Hamiltonian $H$ corresponding to the product of the number of atoms and orbitals considered in the simulation. $\theta_i$ is a random number chosen from $[0,2\pi)$. The error in this approximation is of $\mathcal{O}(1/\sqrt{RD})$, therefore when increasing either $R$ or the number of unit cells within the system allows for rapid convergence with only a few random vectors. 

\section{Expansion of the Kubo bastin formula using Chebysehv polynomials}
Following the presentation of the KPM, we can proceed to expand the orbital-Hall conductivity tensor. Starting from the Kubo-Bastin formula, the components of the orbital-Hall conductivity tensor are given by

\begin{align}
	\sigma_{\alpha\beta}^{k}(\mu,T) & = \frac{ie\hbar}{\Omega}\int_{-\infty}^{\infty}d\varepsilon F(\varepsilon,\mu,T) \text{Tr}\left\langle J_{\alpha}^{k}\partial _{\varepsilon}G^{+}(\varepsilon,H)v_{\beta}\delta(H-\varepsilon) -J_{\alpha}^k\delta(H-\varepsilon)v_{\beta}\partial_{\varepsilon}G^{-}(\varepsilon,H)\right\rangle\nonumber \\
	{} & =  \frac{ie\hbar}{\Omega}\int_{-E_{min}}^{E_{max}}d\varepsilon F(\varepsilon,\mu,T) \text{Tr}\left\langle J_{\alpha}^{k}\partial _{\varepsilon}G^{+}(\varepsilon,H)v_{\beta}\delta(H-\varepsilon) -J_{\alpha}^k\delta(H-\varepsilon)v_{\beta}\partial_{\varepsilon}G^{-}(\varepsilon,H)\right\rangle.
	\label{eqn:fullkubobastinOHE}
\end{align}


\noindent Here, $\Omega$ is the volume of the sample, $G^{\pm}(\varepsilon,H)$ are the retarded (advanced) Green's functions, $v_{\beta}$ is the velocity operator in the $\beta$ direction and $F(\varepsilon)$ is the Fermi-Dirac function $F(\varepsilon,\mu,T)=(\exp((\varepsilon-\mu)/K_BT)+1)^{-1}$, and $J_{\alpha}^{k}=\frac{1}{2}\lbrace L_{k},v_{\alpha} \rbrace$ is the orbital current operator. In \eqref{eqn:fullkubobastinOHE}, we restricted the integration interval under the assumption that the Hamiltonian possesses finite lower and upper bounds, which is always true for tight-binding Hamiltonians. Rescaling the Hamiltonian $H$, the energies $\varepsilon$, the chemical potential $\mu$, and the temperature $K_BT$ we get:

\begin{align}
	\sigma_{\alpha\beta}^{k}(\mu,T) &= \left(\frac{2}{E_{max}-E_{min}}\right)^2 \frac{ie\hbar}{\Omega}\int_{-1}^{1}d\tep F(\tep,\tilde{\mu},\tilde{T})\text{Tr}\left\langle J_{\alpha}^{k} \partial_{\tep}G^{+}(\tep,\tH)v_{\beta}\delta(\tep-\tH) - J_{\alpha}^{k}\delta(\tep - \tH)v_{\beta}\partial_{\tep}G^{-}(\tep,\tH) \right\rangle\nonumber\\
	{ }&= \left(\frac{2}{E_{max}-E_{min}}\right)^2 \tilde{\sigma}_{\alpha\beta}^{k}(\tilde{\mu},\tilde{T})
	\label{eqn:scaledkubo}
\end{align}

From now on, we can work with the rescaled orbital-Hall conductivity tensor $\tilde{\sigma}_{\alpha\beta}^{k}(\tilde{\mu},\tilde{T})$. Using \eqref{eqn:chebexpcoef}, we can expand the $\delta(\varepsilon-H)$ and $G^{\pm}(\varepsilon,H)$ in terms of Chebyshev polynomials as:

\begin{equation}
	\delta(\tep-\tH) = \frac{2}{\pi\sqrt{1-\tep^2}}\sum_{m=0}^{M}g_mT_{m}(\tep)T_{m}(\tH),\qquad G^{\pm} =\mp\frac{i}{\sqrt{1-\tep^2}}\sum_{m=0}^{M}(2-\delta_{m0})g_m\exp(\mp im\arccos(\tep))
\end{equation}

replacing this in \eqref{eqn:scaledkubo}, we get

\begin{equation}
	\tilde{\sigma}_{\alpha\beta}^{k} = -\frac{4 e  \hbar}{\pi\Omega}\int_{-1}^{1}d\tep\frac{f(\tep,\tilde{\mu},\tilde{T})}{(1-\tep^2)^2}\sum_{mn}\mu_{mn}^{\alpha\beta k}2\text{Im}\left(\Gamma_{mn}(\tep) \right),
	\label{eqn:fullcompressedChebKubo}
\end{equation}

where the expansion moments are

\begin{equation}
	\mu_{mn}^{\alpha\beta k} = \frac{g_m g_n}{(1+\delta_{m0})(1+\delta_{n0})}\text{Tr}\langle J_{\alpha}^{k}T_m(\tH)v_{\beta}T_n(\tH)\rangle, 
\end{equation}

\noindent they do not depend on $\tep$, which carries all the information of the system and their computation is responsible for most of the computational cost. On the other hand,

\begin{equation}
	\Gamma_{mn}(\tep) = \left(-i\tep +m\sqrt{1-\tep^2}\right)T_n(\tep)\exp(-im\arccos(\tep))
\end{equation}

\noindent is independent of the Hamiltonian and can be interpreted as the expansion basis. Therefore, \eqref{eqn:fullcompressedChebKubo} can be seen as a generalization of \eqref{eqn:DOSexpansion} where more than two spectral functions are present in the expansion. Moreover, using the trace properties one can show that $\mu_{mn}^{\alpha\beta k}$ obey the symmetry

\begin{equation}
	\left(\mu_{mn}^{\alpha\beta k}\right)^{\ast} =\mu_{nm}^{\alpha\beta k}\label{eqn:momsymm}
\end{equation}

\section{Computation of the orbital angular momentum operator and the expansion moments}

In the previous section, we outlined the general procedure for expanding the orbital-Hall conductivity tensor computed from the Kubo-Bastin formula. It is general for any other operator beyond the orbital current operator. This section will focus on the details of the computation of $L_k$ and the coefficients $\mu_{mn}^{\alpha\beta k}$. First, we recall the definition of the orbital moment 

\begin{equation}
	L_{k} = \epsilon_{ijk}\frac{e\hbar^2}{4g_L \mu_B}\left(\frac{r_{i}^{\mathbf{+}}}{\hbar}v_{j}- v_{i}\frac{r_{j}^{\mathbf{-}}}{\hbar}\right),\label{eqn:orbitalmomentoperator}
\end{equation}

\noindent where Einstein's summation convention is assumed, and latin indexes were used to differentiate the operators used for the composition of the orbital moment $L_k$ from the velocity operators related to the electric field and orbital current definitions within the Kubo-Bastin formula \eqref{eqn:fullkubobastinOHE}. Recalling the results from Equation 2 of the manuscript, the off-diagonal elements of the position operators $r_i^{+}$ and $r_i^{-}$ are

\begin{align}
	\langle a|r_{i}^{+}|b\rangle =i\frac{\hbar}{2}\langle a|\int_{E_{min}}^{E_{max}} d\varepsilon^{\prime}\left[(G^{+}(\varepsilon^{\prime},H)+G^{-}(\varepsilon^{\prime},H))v_{i}\delta(H-\varepsilon^{\prime})\right]|b\rangle=i\frac{\hbar}{2}\lim_{s\rightarrow0^+}\left(\frac{\langle a |v_{i}|b\rangle}{(\varepsilon_b -\varepsilon_a + i s)} + \frac{\langle a |v_{i}|b\rangle}{(\varepsilon_b -\varepsilon_a - i s)}\right)\nonumber&&\\
	\langle a|r_{i}^{-}|b\rangle  =-i\frac{\hbar}{2}\langle a|\int_{E_{min}}^{E_{max}}d\varepsilon^{\prime}\left[\delta(H-\varepsilon^{\prime})v_{i}(G^{-}(\varepsilon^{\prime},H)+G^{+}(\varepsilon^{\prime},H))\right]|b\rangle= -i\frac{\hbar}{2}\lim_{s\rightarrow0^+}\left(\frac{\langle a |v_{i}|b\rangle}{(\varepsilon_a -\varepsilon_b - i s)} + \frac{\langle a |v_{i}|b\rangle}{(\varepsilon_a -\varepsilon_b + i s)} \right)\label{eq:positionoperators}
\end{align}

\noindent Where $|a\rangle$ and $|b\rangle$ are the eigenstates of the Hamiltonian $H$ with energies given by $H|a\rangle = \varepsilon_{a}|a\rangle$ and $H|b\rangle = \varepsilon_{b}|b\rangle$, respectively. As mentioned in the main text, the operators in \eqref{eq:positionoperators} are equivalent to the off-diagonal elements of the position operator $\langle i| r_{\alpha}|j\rangle$ in the limit of vanishing broadening. Following the procedure outlined in equation \eqref{eqn:chebexpspectral}, the energy integrals can be discretized and written in terms of their Chebyshev polynomial expansions as

\begin{align}
	\langle a|r_{i}^{+}|b\rangle = i\hbar\frac{2}{\Delta E}\sum_{p=0}^{2M-1} \Delta \tep \langle a |\left(\sum_{\mu=0}^{M}\text{Re}(c_{\mu}^{+}(\tep_p))T_{\mu}(\tH)\right) v_{i}\left(\sum_{\nu=0}^{M}f_{\nu}(\tep_p)T_{\nu}(\tH)\right)|b\rangle \nonumber&&\\
	\langle a|r_{i}^{-}|b\rangle = -i\hbar\frac{2}{\Delta E}\sum_{p=0}^{2M-1} \Delta\tep\langle a |\left(\sum_{\mu=0}^{M}f_{\mu}(\tep_p)T_{\mu}(\tH)\right) v_{i}\left(\sum_{\nu=0}^{M}\text{Re}(c_{\nu}^{-}(\tilde{\varepsilon}_p))T_{\nu}(\tH)\right)|b\rangle.\label{eqn:opexpansion}
\end{align}

In \eqref{eqn:opexpansion}, we have discretized the energy integral, considered $2M$ points for the numerical integration, used $\eta \approx 0.99$, and defined $\Delta\tep=\frac{2\times\eta}{2M-1}$  and $\tep_p = -\eta+p\times \Delta\tep$. This integration range was chosen to avoid numerical instabilities due to the divergence of the Chebyshev polynomials at $\tilde{\varepsilon}=\pm1$. The expansion coefficients are:

\begin{equation}
	c_{\mu}^{\pm}(\tep)=\frac{\mp 2i}{\sqrt{1-\tep^2}}\frac{g_{\mu}e^{\mp i\mu\arccos(\tep)}}{(\delta_{\mu,0}+1)} \qquad\text{and} \qquad f_{\nu}(\tep)=\frac{2}{\pi \sqrt{1-\tep^2}}\frac{g_{\nu}T_{\nu}(\tep)}{(\delta_{\mu,0}+1)},
\end{equation}

\noindent and correspond to the expansion of $G^{\pm}(\varepsilon,H)$ and $\delta(H-\varepsilon)$, respectively\cite{GarciaRappoportConductivity}. Finally, the coefficients of the expansion of the orbital Hall conductivity tensor $\mu_{mn}^{\alpha\beta k}$  are

\begin{align}
	\mu_{mn}^{\alpha\beta k} &= \frac{g_m g_n}{2\left(1+\delta_m,0\right)\left(1+\delta_n,0\right)} Tr\langle \lbrace \ell_{k},v_{\alpha}\rbrace T_{m}(\tilde{H})v_{\beta}T_{n}(\tilde{H})\rangle\nonumber\\
	{ }&=\epsilon_{ijk}\frac{g_m g_n}{2\left(1+\delta_m,0\right)\left(1+\delta_n,0\right)} \left[\underbrace{Tr\langle \left(\frac{r^{+}_i}{\hbar}v_j -v_i\frac{r^{-}_j}{\hbar}\right) v_{\alpha} T_m(\tilde{H})v_{\beta}T_n(\tilde{H})\rangle}_{A} + \overbrace{Tr\langle v_{\alpha}\left(\frac{r^{+}_i}{\hbar}v_j -v_i\frac{r^{-}_j}{\hbar}\right)T_m(\tilde{H})v_{\beta}T_n(\tilde{H})\rangle}^{B}\right]. \label{eqn:Moment1}
\end{align}

\noindent Here, we define $\ell_k=\frac{1}{\hbar}L_k$ to group the computation of $L_k$ without constants. In \eqref{eqn:Moment1}, we have labeled the two terms $A$ and $B$ as the elements resulting from the anticommutator in the definition of the orbital currents. Labeling the elements depending on each of the position operators in \eqref{eq:positionoperators} $A$ reads as

\begin{align}
	A=\overbrace{ Tr\langle\frac{r^{+}_i}{\hbar}v_j v_\alpha T_m(\tilde{H})v_{\beta}T_n(\tilde{H})\rangle }^{A^{+}} - \overbrace{Tr\langle v_i\frac{r^{-}_j}{\hbar}v_\alpha T_m(\tilde{H})v_{\beta}T_n(\tilde{H})\rangle}^{A^{-}}.
\end{align}

Inserting the expansions in \eqref{eqn:opexpansion}, we get:

\begin{align}
	A^{+}=\frac{2}{E_{max}-E_{min}} \sum_{p=0}^{2M-1} \left[i\Delta \tilde{\varepsilon}\langle\Psi|\left(\sum_{\mu=0}^{M}\text{Re}(c_{\mu}^{+}(\tilde{\varepsilon}^{\prime}_p))T_{\mu}(H)\right) v_{i}\left(\sum_{\nu=0}^{M}f_{\nu}(\tilde{\varepsilon}^{\prime}_p)T_{\nu}(\tilde{H})\right)v_j\right]v_{\alpha}T_m(\tilde{H})v_{\beta}T_n(\tilde{H})|\Psi\rangle\nonumber\\
	A^{-}=\frac{2}{E_{max}-E_{min}} \sum_{p=0}^{2M-1} \left[-i\Delta \tilde{\varepsilon}\langle\Psi|\left(\sum_{\mu=0}^{M}f_{\mu}(\tilde{\varepsilon}^{\prime}_p)T_{\mu}(H)\right) v_{i}\left(\sum_{\nu=0}^{M}\text{Re}(c_{\nu}^{-}(\tilde{\varepsilon}^{\prime}_p))T_{\nu}(\tilde{H})\right)v_j\right]v_{\alpha}T_m(\tilde{H})v_{\beta}T_n(\tilde{H})|\Psi\rangle,\label{eq:FinalMoments1}
\end{align}

In \eqref{eq:FinalMoments1}, we take advantage of our approach and apply the complex position operators defined as the numerical integrals in \eqref{eqn:opexpansion} over the $\langle\Psi|$. Replicating the same procedure for $B$, we cancel out the spurious contributions related to the imaginary part from the definition of \eqref{eq:positionoperators} using the symmetry of the expansion moments shown in \eqref{eqn:momsymm}. Therefore, we define the final coefficients as 

\begin{equation}
	\left(\frac{2}{E_{max}-E_{min}}\right)\Phi_{m,n}^{\alpha\beta k}=\frac{1}{2}\left(\mu_{m,n}^{\alpha\beta k} + (\mu_{n,m}^{\alpha\beta k})^{\ast}\right),\label{eqn:finalcoef}
\end{equation}

\noindent here we collected all the prefactors and symmetrized the expansion coefficients. Finally, substituting \eqref{eqn:finalcoef} in \eqref{eqn:fexpansion}, we arrive to the equation 4 of the main text

\begin{align}
	\sigma_{\alpha\beta}^{k} &= -\left(\frac{2}{E_{max}-E_{min}}\right)^{3} \frac{e^2 \hbar^3}{4g_L\mu_B\Omega} \frac{4}{\pi}\int_{-1}^{1}d\tep \frac{F(\tep,\tilde{\mu},\tilde{T})}{(1-{\tep}^2)^2}\sum_{m,n=0}^{M} 2 \Phi_{mn}^{\alpha,\beta k}\text{Im}(\Gamma_{mn}(\tep)),\label{eqn:fexpansion}
\end{align}

\section{Energy dependent computation of the orbital moment}

To show the equivalence between the well-known expression for the orbital moment of Bloch electrons and our alternative representation, we will compare the energy-dependent orbital moment obtained through our method and the orbital moment obtained from the modern theory of orbital magnetization in the topologically non-trivial phase of the Haldane model. For this, let us recall the expression for the orbital moment used in Ref. \cite{BuschMertig}, where the orbital moment is given by 

\begin{align}
	\langle \nu \bm{k}| m_z |\alpha \bm{k}\rangle =& i \frac{e\hbar}{4}\sum_{\beta \neq \nu,\alpha}\left(\frac{1}{\varepsilon_{\beta\bm{k}}-\varepsilon_{\nu\bm{k}}}+\frac{1}{\varepsilon_{\beta\bm{k}}-\varepsilon_{\alpha\bm{k}}}\right) \nonumber\\
	{ } & \times \left(\langle \nu\bm{k}|v_x|\beta\bm{k}\rangle \langle \beta \bm{k}|v_y|\alpha\bm{k}\rangle - \langle\nu\bm{k}|v_y|\beta\bm{k}\rangle \langle \beta \bm{k}|v_x|\alpha\bm{k}\rangle\right).
\end{align}

\begin{figure}[ht]
	\centering
	\includegraphics[width=0.75\linewidth]{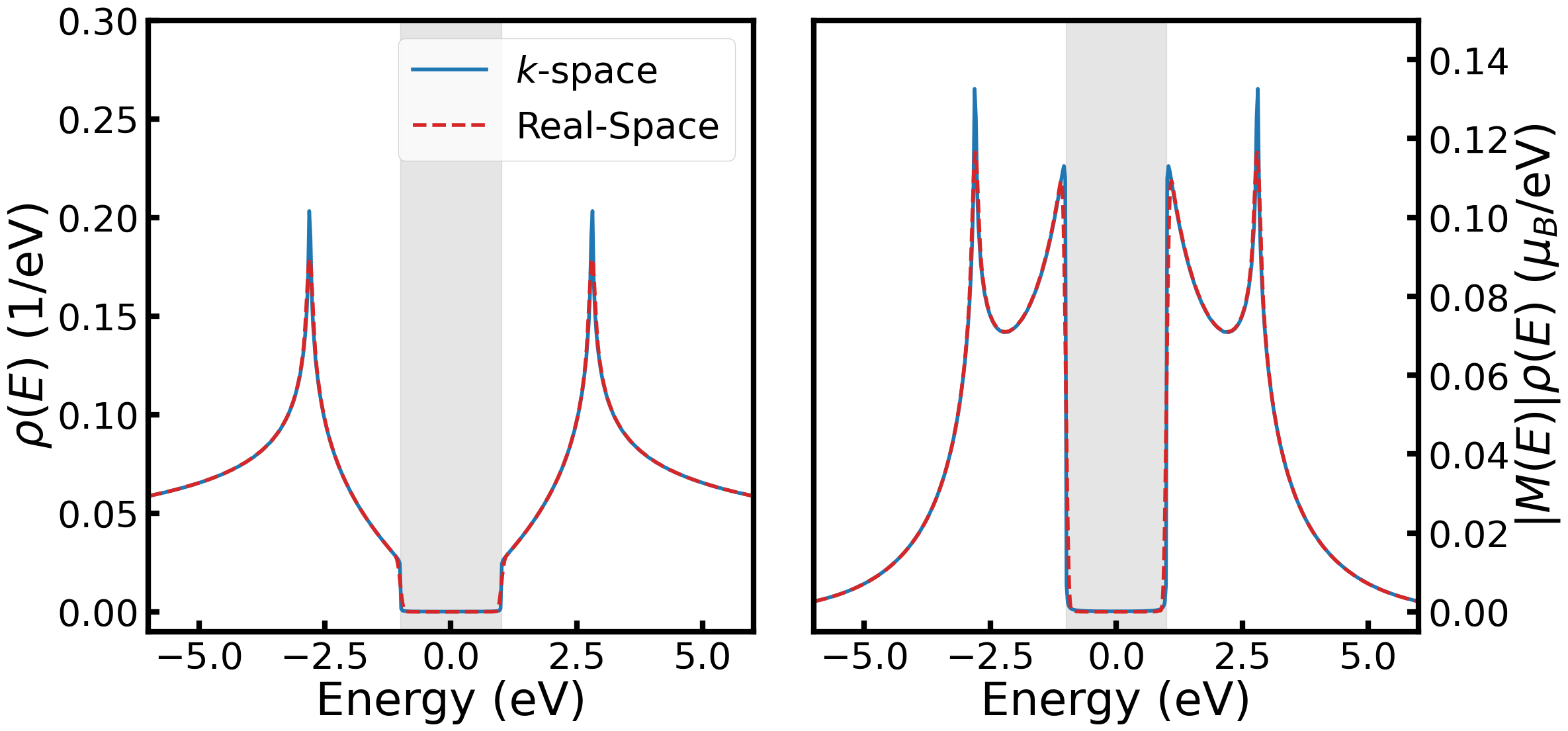}
	\caption{{\bf{Left:}} Density of states for the Haldane model in the topologically nontrivial phase with $t_2= 1.0$ eV and $\Delta_i=0$, computed using the Green's function method in momentum space (blue solid line) and with the kernel polynomial method (red dashed lines). {\bf{Right:}} Energy-dependent orbital moment for the Haldane model in the topologically nontrivial phase with $t_2=1.0$ eV and $\Delta_i=0$, computed using the Green's function method in momentum space (blue solid line) and with the kernel polynomial method (red dashed lines). The shaded area signals the energy gap. For our Green's function calculations, we used a Monkhorst-Pack grid of $N_{\bm{k}}=4096\times4096$ points with broadening $\eta=25\times10^{-4}$ eV, while for our KPM calculations, we used a real-space system composed of $512\times512$ unit cells and $M=1024$ Chebyshev polynomials for a numerical broadening of  $\delta  \varepsilon\approx 0.44$ eV. }
	\label{fig:OMCompare}
\end{figure}

Since the Haldane model solely contains two bands, the restriction in the summation imposes that the orbital moment matrix has only non-zero elements in the diagonal. Therefore, we can evaluate directly these matrix elements as 

\begin{align}
	\langle 1\bm{k}|m_z|1\bm{k}\rangle = & -\frac{e\hbar}{4}\left(\frac{4}{\varepsilon_{2\bm{k}}-\varepsilon_{1\bm{k}}}\right)\text{Im}\left(\langle 1\bm{k}|v_x|2\bm{k}\rangle \langle 2 \bm{k}|v_y|1\bm{k}\rangle\right)\nonumber\\
	\langle 2\bm{k}|m_z|2\bm{k}\rangle = & -\frac{e\hbar}{4}\left(\frac{4}{\varepsilon_{1\bm{k}}-\varepsilon_{2\bm{k}}}\right)\text{Im}\left(\langle 2\bm{k}|v_x|1\bm{k}\rangle \langle 1 \bm{k}|v_y|2\bm{k}\rangle\right),
	\label{eqn:OMgrapheneKspace}
\end{align}

\noindent where $|1\bm{k}\rangle$ and $|2\bm{k\rangle}$ correspond to the eigenstates of the conduction and valence band, respectively, at the momentum $\bm{k}$. Manipulating \eqref{eqn:OMgrapheneKspace}, one can find that both energy bands have the same orbital moment for a given $\bm{k}$. To complete our derivation, we obtain the energy-dependent orbital moment from the product of the orbital moment matrix $m_z(\bm{k})$ with the density of states $\displaystyle\delta(H(\bm{k})-\varepsilon)= \mp \frac{1}{\pi} \text{Im}(G^{\pm}(\varepsilon,H(\bm{k})))$ to get the energy-dependent orbital moment in the system. Therefore, the energy-dependent orbital moment reads like 
\begin{equation}
	\langle m_z(\varepsilon)\rangle = -\frac{1}{2N_{\bm{k}}\pi}\sum_{\bm{k} \in BZ}\langle\bm{k}| m_z(\bm{k}) \text{Im}(G^{+}(\varepsilon,H(\bm{k}))) - \text{Im}(G^{-}(\varepsilon,H(\bm{k}))) m_z(\bm{k})|\bm{k} \rangle, 
\end{equation}
where $N_{\bm{k}}=4096\times4096$ is the number of points used in the Monkhorst-Pack sampling of the Brillouin zone. In our Green's function calculations, we used a broadening $\eta=25\times10^{-4}$ eV. To compare with our method, we express the energy-dependent orbital moment as 

\begin{equation}
	\langle m_z(\varepsilon)\rangle = \sum_{r=0}^{R-1}\frac{1}{2 R}\langle \psi_r|m_z\delta(\varepsilon-H)+ \delta(\varepsilon-H)m_z|\psi_r\rangle
\end{equation}

where, we used $\displaystyle m_z=\frac{g_L\mu_B}{\hbar}L_z$ and $R$ is the total number of random vectors. In our real-space calculations, we considered a lattice composed of $512\times512$ unit cells of the Haldane model and $M=1024$ Chebyshev polynomials in our KPM expansion for a broadening  $\delta  \varepsilon\approx 0.44$ eV. 
From Fig. \ref{fig:OMCompare}, it can be seen that our method yields similar results as the orbital moment computed with \eqref{eqn:OMgrapheneKspace} with the only differences being the slight broadening of the peaks at the gap edge $E=\pm1.0$ eV and at the van Hove singularities $E=\pm2.8$ eV due to the numerical broadening introduced in the Chebyshev polynomial expansion by the kernel that dampens the Gibbs oscillations.

\clearpage

\section{Energy dependent orbital Hall conductivity for gapped graphene}

\begin{figure}[ht]
	\centering
	\includegraphics[width=0.75\linewidth]{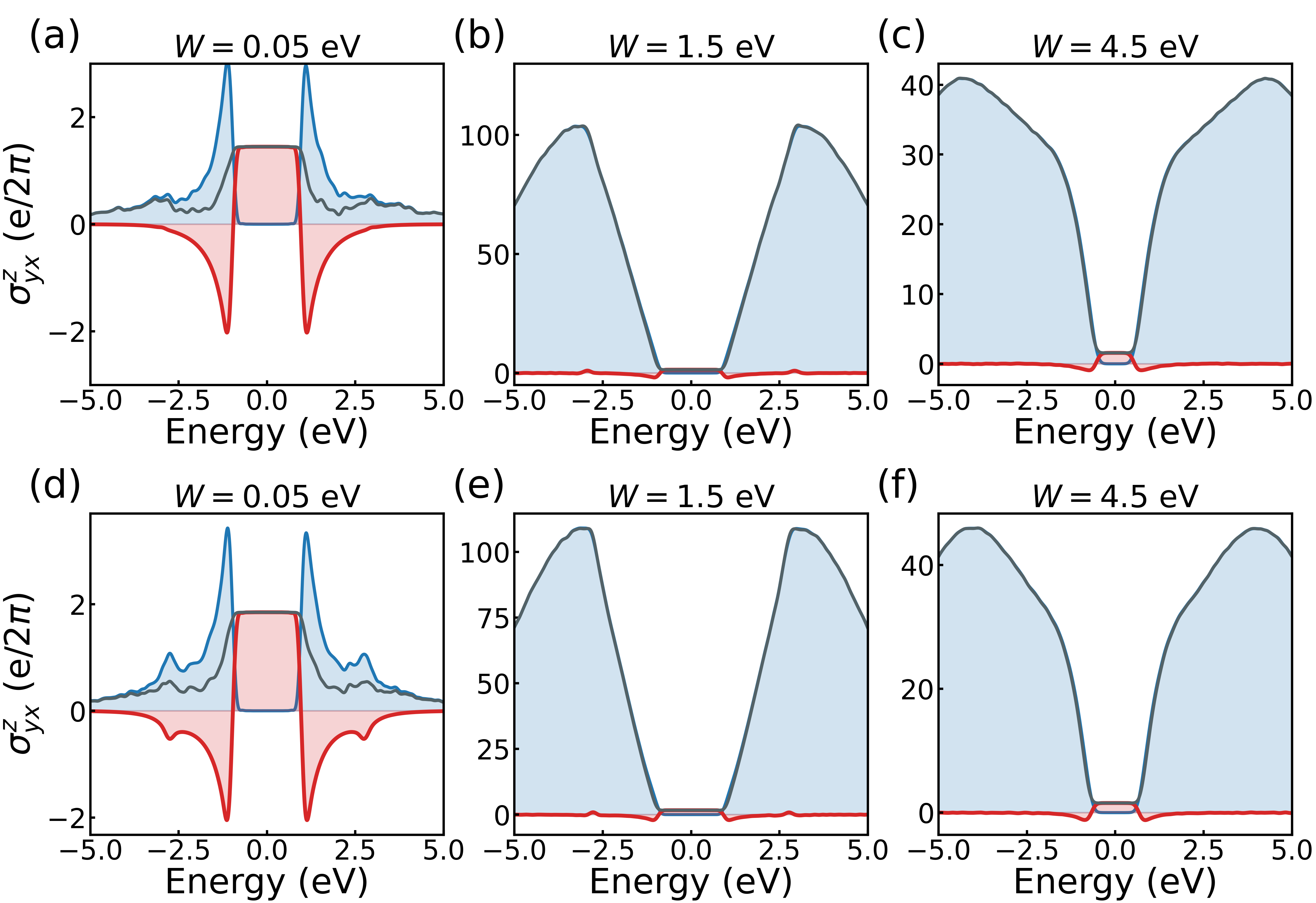}
	\caption{Fermi sea (red shaded area) and Fermi surface (blue shaded area) contributions to the total orbital Hall conductivity (grey line) for the Haldane model in the trivial phase for $t_2=0$ (upper row), and the topologically nontrivial phase for $t_2=1.0$ eV, $\Delta=0.0$ (lower row) computed for systems with $512\times512$ unit cells, $M=512$  and averaged over $160$ disorder configurations.}
	\label{fig:DisorderDep}
\end{figure}

In Figure 3 of the main text, we showed the variation of the sea contributions to the OHC as a function of the Anderson disorder strength $W$. Fig. \ref{fig:DisorderDep} shows the energy-dependent data for some of these points to illustrate variations of the sea and surface contributions with $W$.

\clearpage

\begin{figure}[ht]
	\centering
	\includegraphics[width=0.6\linewidth]{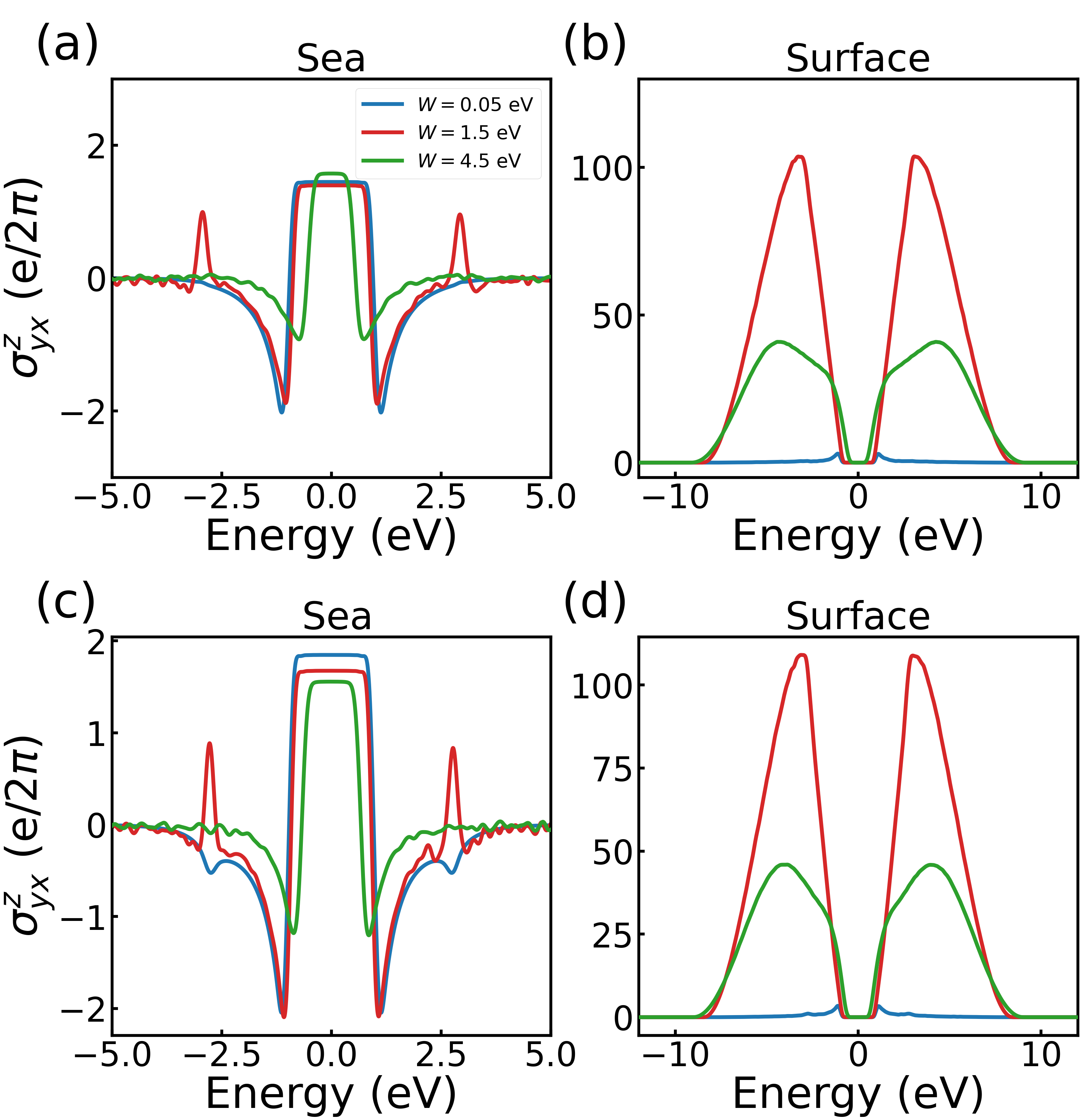}
	\caption{Change in the Fermi sea ((a) and (c)) Fermi surface ((b) and (d)) contributions of the orbital Hall conductivity for the Haldane model for various disorder values $W$ in the trivial phase for $t_2=0$ (upper row), and the topologically nontrivial phase for $t_2=1.0$ eV, $\Delta=0.0$ (lower row) computed for systems with $512\times512$ unit cells, $M=512$  and averaged over $160$ disorder configurations.}
	\label{fig:DisorderStreda}
\end{figure}

To illustrate better the changes in the contributions with the disorder, in Fig. \ref{fig:DisorderStreda}, we show the sea and surface components for three values of Anderson disorder. Upon inspection, it is clear that the surface contributions dominate the intrinsic ones. However, the topological contributions are mostly unchanged with small increments for $W=4.5$ eV (decrements) for the topologically trivial (non-trivial) phases.

\end{document}